\documentclass[twocolumn,authoryear]{autart}    
\usepackage{mathrsfs}
\usepackage{amsfonts}
\usepackage{bbm}
\usepackage{amsmath,amssymb}
\usepackage{graphicx}
\usepackage{subfigure}
\usepackage{epstopdf}
\usepackage{subfigure}
\usepackage{epic}
\usepackage{tikz}
\usepackage{algorithm}
\usepackage{algpseudocode}
\usepackage{graphics} 
\usepackage{color}
\usepackage[justification=centering]{caption}
\usetikzlibrary{arrows,shapes,chains}
\usepackage{cite}
\usepackage{algorithm}
\usepackage{algpseudocode}
\allowdisplaybreaks

\newtheorem{definition}{Definition}[section]
\newtheorem{theorem}{Theorem}[section]
\newtheorem{lemma}{Lemma}[section]

\newtheorem{remark}{Remark}[section]
\newtheorem{assumption}{Assumption}[section]

\newtheorem{example}{Example}[section]

\begin{document}

\begin{frontmatter}

\title{Robust approximate symbolic models for a class of continuous-time uncertain nonlinear systems\\
 via a control interface\thanksref{footnoteinfo}} 

\thanks[footnoteinfo]{This work was supported in part by the Swedish Research Council (VR), the European Research Council (ERC), the Swedish Foundation for Strategic Research (SSF) and the Knut and Alice Wallenberg Foundation (KAW). \\
Corresponding author. Tel.: +46 7 64199298.
E-mail address: piany@kth.se (Pian Yu).}
\author[KTH]{Pian Yu},    
\author[KTH]{Dimos V. Dimarogonas}

\address[KTH]{Division of Decision and Control Systems, EECS, KTH Royal Institute of Technology, 10044 Stockholm, Sweden.}             


\begin{abstract}                          
Discrete abstractions have become a standard approach to assist control synthesis under complex specifications. Most techniques for the construction of a discrete abstraction for a continuous-time system require time-space discretization of the concrete system, which constitutes property satisfaction for the continuous-time system non-trivial. In this work, we aim at relaxing this requirement by introducing a control interface. Firstly, we connect the continuous-time uncertain concrete system with its discrete deterministic state-space abstraction with a control interface. Then, a novel stability notion called $\eta$-approximate controlled globally practically stable, and a new simulation relation called robust approximate simulation relation are proposed. It is shown that the uncertain concrete system, under the condition that there exists an admissible control interface such that the augmented system (composed of the concrete system and its abstraction) can be made $\eta$-approximate controlled globally practically stable, robustly approximately simulates its discrete abstraction. The effectiveness of the proposed results is illustrated by two simulation examples.
\end{abstract}

\end{frontmatter}

\section{Introduction}

In recent years, discrete abstractions have become one of the standard approaches for control synthesis in the context of complex dynamical systems and specifications \cite{Tabuada09}. It allows one to leverage computational tools developed for discrete-event systems \cite{Ramadge87,Kumar95,Cassandras99} and games on automata \cite{Arnold03,Madhusudan03} to assist control synthesis for specifications difficult to enforce with conventional control design methods, such as linear temporal logic \cite{Baier2008} specifications. Moreover, if the behaviors of the original system (referred to as the concrete system) and the abstract system (obtained by, e.g., discretizing the  state-space) can be formally related by an inclusion or equivalence relation, the synthesized controller is known to be correct by design \cite{Girard12}.

For a long time, (bi)simulation relations were a central notion to deal with complexity reduction \cite{Milner89,Park81}. It was later pointed out in \cite{Alur00} that this kind of equivalence relation is often too strong. To this end, a new notion called approximate (bi)simulation, which only asks for the closeness of observed behaviors, was introduced in \cite{Girard_Pappas07}. Based on the notion of incrementally (input-to-state) stable \cite{Angeli02}, approximately bisimilar symbolic models were built and extended to various systems \cite{Girard_Pola10,Zamani14}. However, incrementally (input-to-state) stable is a strong property for dynamical control systems, which makes its applicability restrictive. In \cite{Zamani12}, the authors relax this requirement by only assuming Lipschitz continuous and incremental forward completeness, and an approximate alternating simulation relation is established by over-approximating the behavior of the concrete system. However, as recently pointed out in \cite{Reissig17}, this approach may result in a refinement complexity issue. To this end, a new simulation relation, called feedback refinement relation is proposed in \cite{Reissig17}. In addition, for monotone systems, the notion of directed alternating simulation relation is proposed for the construction of symbolic models \cite{Kim2017}.

Although continuous-time systems are extensively studied and various abstraction techniques are proposed in the existing literature, most techniques for the construction of symbolic models require time-space discretization of the continuous-time system, which constitute property satisfaction non-trivial since closeness of the observed behaviors between the concrete system and its abstraction is not guaranteed within neighboring discrete time instants. Recently, different approaches have been proposed in the literature to deal with this \cite{Mallik2018,Liu2016,Saoud2020}. In \cite{Mallik2018}, a disturbance simulation relation is introduced for incrementally input-to-state stable nonlinear systems. In \cite{Liu2016,Saoud2020}, symbolic control approaches are proposed for a class of sample-data nonlinear systems, where property satisfaction of the continuous-time systems is guaranteed by equipping the finite abstractions with certain robustness margins \cite{Liu2016} or assume-guarantee contracts \cite{Saoud2020}. While almost all the results are providing behavioral relationships between a time discretized version of the original system and its symbolic model,
in this paper, we provide for the first time a behavioral relationship between the original continuous-time system and its symbolic model.

This paper investigates the construction of symbolic models for continuous-time uncertain nonlinear systems. It improves upon most of the existing results in two aspects: 1) by not requiring time-space discretization of the concrete system and 2)
by being applicable to more general uncertain nonlinear systems with input constraint. The main contributions are as follows.
\begin{itemize}
\item [i)] We propose a novel stability notion, called $\eta$-approximate controlled globally practically stable. This is a property defined on the augmented system (composed of the concrete system and the abstract system) via an admissible control interface. We show that the abstract system can be constructed without time-space discretization. This is crucial for safety-critical applications, in which it is necessary that the trajectories of the concrete system and the abstract system are close enough at all time instants.
\item [ii)] We define a notion of robust approximate simulation relation. It is shown that for an uncertain concrete system, the abstract system can be constructed such that the concrete system robustly approximately simulates the abstraction.
\item [iii)] For the class of incrementally quadratic nonlinear systems, the systematic construction of the admissible control interfaces and robust approximate symbolic models under bounded input set is provided.
\end{itemize}

The introduction of the control interface is inspired by the hierarchical control framework \cite{Girard_Pappas09,Fu13,Yang17,Smith18(2)}, in which an interface is built between a high dimensional concrete system and a simplified low dimensional abstraction of it. Both the concrete system and the abstract system are continuous in \cite{Girard_Pappas09,Fu13,Yang17,Smith18(2)}. In contrast, in this paper, we propose to build a control interface between the continuous-time concrete system and its discrete state-space abstraction. Moreover, in this paper we consider bounded input set (the input set considered in \cite{Girard_Pappas09,Fu13,Yang17,Smith18(2)} is unbounded), which brings additional difficulty to constructing the interface. Therefore, the results in this paper are essentially novel and improved with respect to the existing work.

A preliminary version of this work was accepted by the 58th IEEE Conference on Decision and Control (CDC 2019) \cite{Pian19}. Here, we expand this preliminary version in three main directions. First, the framework is generalized to include time-varying uncertain nonlinear systems. A new stability notion, called $\eta$-approximate controlled globally practically stable, is proposed. Second, a new simulation relation, called robust approximate simulation relation is proposed to deal with uncertainty. Third, an elaborate motion planning example is added in the simulation section.

The remainder of this paper is organized as follows. In Section 2, notation and preliminaries on system properties are provided. The new stability notion and the construction of symbolic morels are presented in Section 3. In Section 4, an application to incrementally quadratic nonlinear systems is provided. Two illustrative examples are given in Section 5 and Section 6 concludes the paper.

\section{\sc  Preliminaries}

\subsection{Notation}
Let $\mathbb{R}:=(-\infty, \infty)$, $\mathbb{R}_{\ge 0}:=[0, \infty)$, $\mathbb{R}_{> 0}:=(0, \infty)$, $\mathbb{Z}_{> 0}:=\{1,2,\ldots\}$ and $\mathbb{Z}_{\ge 0}:=\{0,1,2,\ldots\}$. Denote $\mathbb{R}^n$ as the $n$-dimensional real vector space, $\mathbb{R}^{n\times m}$ as the $n\times m$ real matrix space. $I_n$ is the identity matrix of order $n$ and $1_n$ is the column vector of order $n$ with all entries equal to one. $0_{n\times m}$ is the $n\times m$ matrix with all elements equal to 0. When there is no ambiguity, we use $0$ to represent a matrix with proper dimensions and all its elements equal to $0$. $[a, b]$ and $[a, b[$ denote closed and right half-open intervals with end points $a$ and $b$. For $x_1\in\mathbb{R}^{n_1}, \ldots, x_m\in\mathbb{R}^{n_m}$, the notation $(x_1, x_2, \ldots, x_m)\in \mathbb{R}^{n_1+n_2+\cdots +n_m}$ stands for $[x_1^T, x_2^T, \ldots, x_m^T]^T$. Let $\left|\lambda\right|$ be the absolute value of a real number $\lambda$, and $\|x\|$ and $\|A\|$ be the Euclidean norm of vector $x$ and matrix $A$, respectively. Given a function $f: \mathbb{R}_{\ge 0}\to \mathbb{R}^n$, the supremum of $f$ is denoted by $\|f\|_\infty$, which is given by $\|f\|_\infty:=\sup\{\|f(t)\|, t\ge 0\}$ and $\|f\|_{[0, \tau)}:=\sup\{\|f(t)\|, t\in [0, \tau)\}$. A function $f$ is called bounded if $\|f\|_\infty<\infty$. Given a set $S$, the interior of $S$ is denoted by $\rm{int}(S)$, the boundary of $S$ is denoted by $F_r(S)$ and the power set of $S$ is denoted by $2^{S}$. Given two sets $S_1, S_2$, the notation $S_1\setminus S_2:=\{x| x\in S_1 \; \wedge \; x\notin S_2\}$ stands for the set difference, where $\wedge$ represents the logic operator AND.

A continuous function $\gamma: \mathbb{R}_{\ge 0}\to \mathbb{R}_{\ge 0}$ is said to belong to class $\mathcal{K}$ if it is strictly increasing and $\gamma(0)=0$; $\gamma$ is said to belong to class $\mathcal{K}_{\infty}$ if $\gamma\in \mathcal{K}$ and $\gamma(r)\to \infty$ as $r\to \infty$. A continuous function $\beta: \mathbb{R}_{\ge 0}\times \mathbb{R}_{\ge 0}\to \mathbb{R}_{\ge 0}$ is said to belong to class $\mathcal{K}\mathcal{L}$ if for each fixed $s$, the map $\beta(r, s)$ belongs to class $\mathcal{K}_{\infty}$ with respect to $r$ and, for each fixed $r$, the map $\beta(r, s)$ is decreasing with respect to $s$ and $\beta(r, s)\to 0$ as $s\to \infty$. For a set $\mathcal{A}\subseteq \mathbb{R}^n$ and any $x\in \mathbb{R}^n$, we denote by, $\rm{d}(x, \mathcal{A})$, the point-to-set distance, defined as $\rm{d}(x, \mathcal{A})=\inf_{y\in \mathcal{A}}\{\|x-y\|\}.$

\subsection{System properties}

Consider a continuous-time uncertain nonlinear system of the form
\begin{equation}\label{cs}
  \Sigma:\left\{\begin{aligned}
  \dot x_1(t)&= f(t, x_1(t), u(t), w(t))\\
  y_1(t)&=h(x_1(t)),
  \end{aligned}\right.
\end{equation}
where $x_1(t)\in \mathbb{R}^{n}, y_1(t)\in \mathbb{R}^{l}, u(t)\in U\subseteq \mathbb{R}^{m}$, $w(t)\in W\subseteq \mathbb{R}^{n_w}$ are the state, output, control input, and external disturbance at time $t$, respectively. The input and disturbance are constrained to sets $U$ and $W$, respectively. We assume that $f: [0, \infty)\times \mathbb{R}^{n}\times U \times \mathbb{R}^{n_w}\to \mathbb{R}^{n}$ is piecewise continuous in $t$, continuous in $x_1, u$ and $w$, and the vector field $f$ is such that for any input in $U$, any disturbance in $W$, and any initial condition $x_1(0)\in\mathbb{R}^{n}$, this differential equation has a unique solution. Throughout the paper, we will refer to $\Sigma$ as the concrete system, that is the system that we actually want to control.

Let
\begin{equation}\label{U1}
  \mathcal{U}=\cup_{\tau\in \mathbb{R}_{>0}\cup\{\infty\}}U^{[0, \tau[}
\end{equation}
be the set of all functions that take their values in $U$ and are defined on intervals
of the form $[0, \tau[$. Similarly, one can define $\mathcal{W}=\cup_{\tau\in \mathbb{R}_{>0}\cup\{\infty\}}W^{[0, \tau[}$. Given an input signal $u\in \mathcal{U}$, we use the notation $\text{dom}(u)$ to represent the domain of $u$.

A curve $\xi: [0, \tau[ \to \mathbb{R}^n$ is said to be a trajectory of $\Sigma$ if there exists an input signal $u\in \mathcal{U}$ and a disturbance signal $w\in \mathcal{W}$ satisfying $\dot{\xi}(t)=f(t, \xi(t), u(t), w(t))$ for almost all $t\in [0, \tau[$. A curve $\zeta: [0, \tau[ \to \mathbb{R}^l$ is said to be an output trajectory of $\Sigma$ if $\zeta(t)=h(\xi(t))$ for almost all $t\in [0, \tau[$, where $\xi$ is a trajectory of $\Sigma$. We use $\xi(\xi_0, u, w, t)$ to denote the trajectory point reached at time $t$ under the input signal $u\in \mathcal{U}$ and the disturbance signal $w\in \mathcal{W}$ from initial state $\xi_0$. When the system (\ref{cs}) is deterministic, i.e., $w(t)\equiv 0$, we denote $\xi(\xi_0, u, t):=\xi(\xi_0, u, 0, t)$ for simplicity. Then, $\xi(\xi_0, u, t)$ is the trajectory of the undisturbed system.

In \cite{Angeli99}, the definition of forward complete (FC) is introduced.

\begin{definition}\cite{Angeli99}\label{Def:FC}
A (deterministic) system is called FC if for every
initial condition $x_0\in \mathbb{R}^{n}$ and every input signal $u\in \mathcal{U}$, the
corresponding solution is defined for all $t\ge 0$.
\end{definition}

By a minor modification of the statement of Definition \ref{Def:FC}, one can define FC for uncertain systems.

\begin{definition}
The uncertain system (\ref{cs}) is called FC if for every
initial condition $x_1(0)\in \mathbb{R}^{n}$, every input signal $u\in \mathcal{U}$, and every disturbance signal $w\in \mathcal{W}$, the
corresponding solution is defined for all $t\ge 0$.
\end{definition}
%

The following definition of $\varepsilon$-closeness characterizes the closeness between two (output) trajectories.

\begin{definition}[\cite{Goedel12}, Definition 4.13]
Given $\varepsilon>0$, two output trajectories $\zeta_1: [0, \infty)\to \mathbb{R}^l$ and $\zeta_2: [0, \infty)\to \mathbb{R}^l$ are $\varepsilon$-close if
\begin{equation*}
  \|\zeta_1(t)-\zeta_2(t)\|\le \varepsilon, \forall t\in [0, \infty).
\end{equation*}
\end{definition}

\begin{lemma}\label{lem1}
Let $V: [0, \infty)\times \mathbb{R}^n\times \mathbb{R}^m \to \mathbb{R}$ be a continuously differentiable function such that
\begin{equation*}
  \begin{aligned}
  \underline\alpha(\|x\|)&\le V(t, x, u)\le \bar\alpha(\|x\|)\\
  \frac{\partial V}{\partial t}+\frac{\partial V}{\partial x}f(t, x, u)&\le -\gamma V(t, x, u), \quad \forall \|x\|\ge \mu >0,
  \end{aligned}
\end{equation*}
$\forall t\ge 0$ and $\forall x\in \mathbb{R}^n$, where $\underline\alpha, \bar\alpha$ are class $\mathcal{K}_\infty$ functions, $\mu>0, \gamma>0$ are constants. Then, the solution $x(t)$ to the differential equation $\dot x=f(t, x, u)$ exists and satisfies
\begin{equation*}
  \|x(t)\|\le \beta(\|x(0)\|, t)+\underline{\alpha}^{-1}(\bar{\alpha}(\mu)),
\end{equation*}
where
\begin{equation*}
\begin{aligned}
\beta(r, t)&=\underline{\alpha}^{-1}(e^{-\gamma t}\bar \alpha(r))
\end{aligned}
\end{equation*}
is a class $\mathcal{K}\mathcal{L}$ function.
\end{lemma}

\textbf{Proof:} The proof follows from Lemma 4.4 and Theorem 4.18 of \cite{Khail02}, and hence omitted. $\hspace*{\fill}\square$

\section{Main results}

\subsection{\sc $\eta$-approximate controlled globally practically stable}

In this paper, the abstraction technique developed in \cite{Girard_Pola10} is applied, in which the state-space $\mathbb{R}^n$ is approximated by the lattice
\begin{equation}\label{sabs0}
  [\mathbb{R}^n]_{\eta}=\Big\{q\in \mathbb{R}^n| q_i=k_i\frac{2\eta}{\sqrt{n}}, k_i\in \mathbb{Z}, i=1, \ldots, n\Big\},
\end{equation}
where $\eta\in \mathbb{R}_{\ge 0}$ is a state-space discretization parameter. Define the associated quantizer $Q_\eta: \mathbb{R}^n\to [\mathbb{R}^n]_{\eta}$ as $Q_\eta(x)=q$ if and only if $|x_i-q_i|\le \eta/\sqrt{n}, \forall i=1, \ldots n$. Then, one has $\|x-Q_\eta(x)\|\le \eta, \forall x\in \mathbb{R}^n$.

The abstract system is obtained by applying the state abstraction (\ref{sabs0}) to the undisturbed concrete system, which is given by
\begin{equation}\label{abs}
  \Sigma':\left\{\begin{aligned}
  x_2(t)&= Q_\eta(\hat x_2(t)),\\
  \dot{\hat x}_2(t)&=f_d(t, \hat x_2(t), v(t)),\\
  y_2(t)&=h(x_2(t)),
  \end{aligned}\right.
\end{equation}
where $x_2(t)\in [\mathbb{R}^n]_\eta, \hat x_2(t)\in \mathbb{R}^n, y_2(t)\in \mathbb{R}^l$, and $v(t)\in U'(t)$ represent the state, output and control input of the abstract system, respectively. Initially, ${\hat x}_2(0)={x}_2(0)$. The function $f_d: [0, \infty)\times \mathbb{R}^{n}\times \mathbb{R}^m \to \mathbb{R}^{n}$ represents the nominal dynamics of the concrete system (\ref{cs}), i.e., $f_d(t, x, u)= f(t, x, u, w)$ if $w=0$. Note that the state variable $x_2(t)$ is neither continuous nor differentiable due to the state-space discretization. In addition, we note that the map $U': \mathbb{R}_{\ge 0}\to 2^{\mathbb{R}^m}$ denotes the possibly time-varying input constraint for the abstract system, which is a design parameter that will be specified later.

Let $R_{U'}:=\{U'(t): t\in \mathbb{R}_{\ge 0}\}$ be the range of the set-valued function $U'$. Then, we define
\begin{equation}\label{U2}
 \mathcal{U}'=\cup_{\tau\in \mathbb{R}_{>0}\cup\{\infty\}}R_{U'}^{[0, \tau[}
\end{equation}
as the set of all functions of time from intervals of the form $[0, \tau[$, such that the value of the function at a particular time instant $t$, is an element of $U'(t)$.

A (hybrid) curve $\xi': [0, \tau[ \to [\mathbb{R}^n]_\eta$ is said to be a trajectory of $\Sigma'$ if there exists $v\in \mathcal{U}'$ satisfying $\xi'(t)=Q_\eta(\xi(t)), \forall t\in [0, \tau[,$ where $\dot\xi(t)=f_d(t, \xi(t), v(t))$ and $\xi(0)=\xi'(0)$. A curve $\zeta': [0, \tau[ \to \mathbb{R}^l$ is said to be an output trajectory of $\Sigma'$ if $\zeta'(t)=h(\xi'(t))$, for almost all $t\in [0, \tau[$, where $\xi'$ is a trajectory of $\Sigma'$. With a little abuse of notation, we use $\xi'(\xi'_0, v, t)$ to denote the trajectory point of $\Sigma'$ reached at time $t$ under the input signal $v\in \mathcal{U}'$ from an initial condition $\xi'_0\in [\mathbb{R}^n]_\eta$.

The control input $u(t)$ of the concrete system (\ref{cs}) will be synthesized hierarchically via the abstract system (\ref{abs}) with a control interface $u_v: \mathbb{R}_{\ge 0}\times \mathbb{R}^m \times \mathbb{R}^n\times [\mathbb{R}^n]_\eta \to \mathbb{R}^m$, which is given by
\begin{equation}\label{con_int}
  u(t)=u_v(t, v(t), x_1(t), x_2(t)).
\end{equation}

Define
\begin{equation}\label{hatX0}
  \hat X_0:=\{(x_1, x_2)|x_1\in \mathbb{R}^n, x_2\in [\mathbb{R}^n]_\eta, \|x_1-x_2\|\le \eta\}.
\end{equation}
To guarantee that the synthesized controller $u(t)$ is applicable to the concrete system (\ref{cs}), it is necessary that $u(t)=u_v(t, v(t), x_1(t), x_2(t))\in U, \forall t\in \text{dom}(u)$. Therefore, we propose the following definition.
\begin{definition}\label{def3}
The control interface $u_v: \mathbb{R}_{\ge 0}\times \mathbb{R}^m\times \mathbb{R}^n\times [\mathbb{R}^n]_\eta \to \mathbb{R}^m$ is called \emph{admissible} if there exists an input map $U': \mathbb{R}_{\ge 0}\to 2^{\mathbb{R}^m}$ that satisfies
\begin{itemize}
  \item[i)] $U'(t)\neq \emptyset, \forall t\in \text{dom}(U')$; and
  \item[ii)] $u(t)=u_v(t, v(t), \xi(\xi_0, u_v, w, t), \xi'(\xi'_0, v, t))\in U, \forall t\in \text{dom}(u), \forall (\xi_0, \xi'_0)\in \hat X_0, \forall v\in \mathcal{U}', \forall w\in \mathcal{W}$.
\end{itemize}
In this case, the input map $U'$ is called \emph{admissible} to $u_v$.
\end{definition}

Then, we introduce the following stability notion, which will be used for the construction of symbolic models.

\begin{definition}\label{def5}
Given the concrete system $\Sigma$ in (\ref{cs}) and the abstract system $\Sigma'$ in (\ref{abs}). The system pair $(\Sigma, \Sigma')$ is called \emph{$\eta$-approximate controlled globally practically stable ($\eta$-CGPS)} if it is FC and there exist an admissible control interface $u_v$, a $\mathcal{K}\mathcal{L}$ function $\beta$, and $\mathcal{K}_\infty$ functions $\gamma_1, \gamma_2$ such that $\forall t\in \mathbb{R}_{\ge 0}$, $\forall (x_0, x'_0)\in \hat X_0, \forall v\in \mathcal{U}', \forall w\in \mathcal{W}$, the following condition is satisfied:
\begin{equation*}
\begin{aligned}
\|\xi(x_0, u_v, &w, t)-\xi'(x'_0, v, t)\|\\
 &\le \beta(\|x_0-x'_0\|, t)
 +\gamma_1(\eta)+\gamma_2(\|w\|_\infty).
\end{aligned}
\end{equation*}
Moreover, $u_v$ is called an \emph{interface} for $(\Sigma, \Sigma')$, associated to the \emph{$\eta$-CGPS} property.
\end{definition}

\begin{remark}\label{rem1}
According to Definitions \ref{def3}-\ref{def5}, a general idea on determining the admissible control interface and the associated input map $U'$ can be provided as follows: firstly, ignore the input constraint for the concrete system (\ref{cs}) by assuming that $U=\mathbb{R}^m$ (in this way, any control interface that maps to $\mathbb{R}^m$ is admissible), and find one or several control interfaces $u_v$ such that $(\Sigma, \Sigma')$ is \emph{$\eta$-CGPS}. Secondly, taking the real input set $U$ into account, refine the control interfaces obtained in the previous step in a way that the admissible ones and the associated input maps are kept.
\end{remark}

\begin{remark}
We note that the notion of $\eta$-CGPS defined in Definition \ref{def5} is essentially different from the notion of incrementally input-to-state stable ($\delta$-ISS) given in \cite{Angeli02}, Definition 4.1 or incrementally forward completeness ($\delta$-FC) given in \cite{Zamani12}, Definition 2.4.  Both $\delta$-ISS and $\delta$-FC are properties defined on the concrete system $\Sigma$ while $\eta$-CGPS is a property defined on the system pair $(\Sigma, \Sigma')$. Moreover, for the concrete system that is not $\delta$-ISS or $\delta$-FC, the $\eta$-CGPS property can still hold for the corresponding system pair (as later shown in Section \ref{IQNS}).
\end{remark}
%
%

\begin{remark}
Another difference between $\delta$-FC in \cite{Zamani12} and $\eta$-CGPS is that the $\beta$ function belongs to class $\mathcal{K}_\infty$ in the Definition of $\delta$-FC while class $\mathcal{K}\mathcal{L}$ in Definition \ref{def5}. In \cite{Zamani12}, the state error $\|\xi(x_0, u, t)-\xi(x'_0, u', t)\|$ is not bounded by the initial state error $\|x_0-x'_0\|$ because $\beta(\|x_0-x'_0\|, t)$ can go to infinity as time goes to infinity. This causes the refinement issues. However, it is shown in Definition \ref{def5} that by properly designing the admissible control interface $u_v$, one can upper bound the state error $ \|\xi(x_0, u_v, t)-\xi(x'_0, v, t)\|$ by a $\mathcal{K}\mathcal{L}$ function $\beta(\|x_0-x'_0\|, t)$ (which vanishes as the time goes to infinity) and a constant $\gamma_1(\eta)$ (we consider a deterministic system here for comparison, i.e., $w\equiv 0$). Therefore, our approach has no refinement issues.
\end{remark}

In the following, the Lyapunov function characterization of the stability notion $\eta$-CGPS is proposed, which is motivated by \cite{Girard_Pappas09}.

\begin{definition}\label{def6}
Given the concrete system $\Sigma$ in (\ref{cs}), the abstract system $\Sigma'$ in (\ref{abs}), a smooth function $V: [0, \infty)\times \mathbb{R}^n\times \mathbb{R}^n\to \mathbb{R}_{\ge 0}$, and a control interface $u_v$.
Function $V$ is called a \emph{$\eta$-CGPS Lyapunov function} for $(\Sigma, \Sigma')$ and $u_v$ is the associated control interface if there exist $\mathcal{K}_\infty$ functions $\underline{\alpha}, \bar{\alpha}, \sigma_1, \sigma_2$, and a constant $\mu>0$ such that:
\begin{itemize}
  \item[i)] $\forall x, x'\in \mathbb{R}^n$,
  \begin{equation}\label{lf1}
  \underline{\alpha}(\|x(t)-x'(t)\|)\le V(t, x(t), x'(t))\le \bar{\alpha}(\|x(t)-x'(t)\|);
  \end{equation}
  \item[ii)] $\forall x, x'\in \mathbb{R}^n, \forall v(t)\in U'(t)$, and $\forall w\in \mathcal{W}$,
\begin{equation}\label{lf3}
\begin{aligned}
&\hspace{-0.8cm}\frac{\partial V}{\partial x}\left\{f(t, x(t),u_v(t, v(t), x(t), Q_\eta(x'(t))), w(t))\right\} \\
  &\hspace{2.5cm}+\frac{\partial V}{\partial x'} f_d(t, x'(t), v(t))+\frac{\partial V}{\partial t}\\
 \le& -\mu V(t, x(t), x'(t))+\sigma_1(\eta)+\sigma_2(\|w\|_\infty).
  \end{aligned}
\end{equation}
\end{itemize}
\end{definition}
Then, we can get the following theorem.

\begin{theorem}\label{thm0}
Given the concrete system $\Sigma$ in (\ref{cs}) and the abstract system $\Sigma'$ in (\ref{abs}). If i) $\Sigma$ is FC, ii) there exists a \emph{$\eta$-CGPS Lyapunov function} for $(\Sigma, \Sigma')$ and with $u_v$ being the associated control interface, and iii) $u_v$ is \emph{admissible}, then, $(\Sigma, \Sigma')$ is \emph{$\eta$-CGPS} and $u_v$ is the \emph{interface} for $(\Sigma, \Sigma')$, associated to the \emph{$\eta$-CGPS} property.
\end{theorem}

\textbf{Proof:}
Let $V$ be the $\eta$-CGPS Lyapunov function for $(\Sigma, \Sigma')$ and $u_v$ the associated control interface. Then, one has (\ref{lf3}) holds and thus
\begin{equation*}
\begin{aligned}
 &\frac{\partial V}{\partial x_1} \{f(t, x_1(t), u_v(t, v(t), x_1(t), Q_\eta(\hat x_2(t))), w(t))\}\\
 &\quad\quad\quad\quad\quad\quad\quad\quad+\frac{\partial V}{\partial \xi} f_d(t, \hat x_2(t), v(t)) + \frac{\partial V}{\partial t}\\
  \le& -\mu V(t, x_1(t), \hat x_2(t))+\sigma_1(\eta)+\sigma_2(\|w\|_\infty)\\
  \le& -\frac{\mu}{2} V(t, x_1(t), \hat x_2(t))
\end{aligned}
\end{equation*}
for all $\|x_1(t)-\hat x_2(t)\|\ge \underline{\alpha}^{-1}\left({2\sigma_1(\eta)+2\sigma_2(\|w\|_\infty))}/{\mu}\right).$ According to Lemma \ref{lem1}, one can further have
\begin{equation}\label{x-xi}
\begin{aligned}
  \|x_1(t)-&\hat x_2(t)\|\le \underline{\alpha}^{-1}(e^{-\frac{\mu}{2} t}\bar \alpha(\|x_1(0)-\hat x_2(0)\|))\\
  &+\underline{\alpha}^{-1}(\bar{\alpha}(\underline{\alpha}^{-1}\left({2\sigma_1(\eta)+2\sigma_2(\|w\|_\infty))}/{\mu}\right))).
\end{aligned}
\end{equation}
Moreover, one has from (\ref{abs}) that $\|x_2(t)-\hat x_2(t)\|=\|Q_\eta(\hat x_2(t))-\hat x_2(t)\|\le \eta$. Thus,
\begin{equation*}
\begin{aligned}
  &\|x_1(t)-x_2(t)\|\\
  \le & \|x_1(t)-\hat x_2(t)\|+\|\hat x_2(t)-x_2(t)\|\\
  \le & \underline{\alpha}^{-1}(e^{-\frac{\mu}{2} t}\bar \alpha(\|x_1(0)-x_2(0)\|))\\
  &+\underline{\alpha}^{-1}(\bar{\alpha}(\underline{\alpha}^{-1}\left({2\sigma_1(\eta)+2\sigma_2(\|w\|_\infty))}/{\mu}\right)))+\eta\\
  \le & \underline{\alpha}^{-1}(e^{-\frac{\mu}{2} t}\bar \alpha(\|x_1(0)-x_2(0)\|))+\underline{\alpha}^{-1}(\bar{\alpha}(\underline{\alpha}^{-1}\left(4\sigma_1(\eta)\right)))\\
  &+\eta+\underline{\alpha}^{-1}(\bar{\alpha}(\underline{\alpha}^{-1}\left({4\sigma_2(\|w\|_\infty))}/{\mu}\right))).
\end{aligned}
\end{equation*}
Combining the fact that $u_v$ is admissible, one can conclude that $(\Sigma, \Sigma')$ is $\eta$-CGPS and $u_v$ is the interface for $(\Sigma, \Sigma')$, associated to the $\eta$-CGPS property.
$\hspace*{\fill}\square$

\subsection{Construction of symbolic models}

In this subsection, the construction of symbolic models for the concrete system (\ref{cs}) is considered. Firstly, the notion of robust approximate simulation relation is proposed.

\begin{definition}\label{def8}
Given the concrete system $\Sigma$ in (\ref{cs}) and the abstract system $\Sigma'$ in (\ref{abs}). Let $\varepsilon>0$ be a given precision and $\tilde \varepsilon\ge 0$. We say that \emph{$\Sigma$ robustly approximately simulates $\Sigma'$ with parameters $(\varepsilon, \tilde\varepsilon)$, denoted by $\Sigma'\preceq_{\mathcal{S}}^{(\varepsilon, \tilde\varepsilon)} \Sigma$}, if:
\begin{itemize}
\item[i)] $\forall x_0\in \mathbb{R}^n, \exists x'_0\in [\mathbb{R}^n]_\eta$ such that $(x_0, x'_0)\in \hat X_0$,
\item[ii)] $\forall (x_0, x'_0)\in \hat X_0$, $\forall v\in \mathcal{U}', \exists u\in \mathcal{U}$ such that $\forall t\ge 0$,
    \begin{equation*}
      \|h(\xi(x_0, u, w, t))-h(\xi'(x'_0, v, t))\|\le \varepsilon, \forall w:\|w\|_\infty < \tilde\varepsilon,
    \end{equation*}
\end{itemize}
where $\mathcal{U}, \mathcal{U}'$ and $\hat X_0$ are defined in (\ref{U1}), (\ref{U2}) and (\ref{hatX0}), respectively.
\end{definition}

\begin{remark}\label{rem4}
Item i) of Definition \ref{def8} guarantees that for any initial state $x_0\in \mathbb{R}^n$, one can always find a state $x'_0$ in the abstracted state-space $[\mathbb{R}^n]_\eta$, such that $(x_0, x'_0)\in \hat X_0$. This item is important since for practical implementations, the initial state of the abstract system is determined by the initial state of the concrete system. In addition, item ii) of Definition \ref{def8} guarantees that for every output trajectory $\zeta'$ in the abstract system $\Sigma'$, there exists an output trajectory $\zeta$ in the concrete system $\Sigma$ such that $\zeta'$ and $\zeta$ are $\varepsilon$-close (despite the worst disturbance signals). Therefore, for a given specification, e.g., a safety and reachability specification $S$, if one can find an output trajectory in $\Sigma'$ such that $S'$ ($S'$ is obtained by enlarging all the unsafe sets by $\varepsilon$ and shrinking all the target sets by $\varepsilon$) is satisfied, then one can always find an output trajectory in $\Sigma$ such that $S$ is satisfied under all possible disturbances.
\end{remark}

\begin{remark}
We note that the robust approximate simulation relation defined in Definition \ref{def8} resembles the notion of disturbance (bi)simulation relation given in \cite{Mallik2018}, Definition 2. The difference lies in that our relation is defined for continuous-time systems while in \cite{Mallik2018}, the relation is defined for discrete-time systems.
\end{remark}

Before proceeding, we need the following additional assumption.

\begin{assumption}\label{ass1}
The output function $h: \mathbb{R}^n\to \mathbb{R}^l$ is globally Lipschitz continuous with Lipschitz constant $\rho$ on the set $X_\varepsilon$. That is,
\begin{equation*}
  \|h(x_1)-h(x_2)\|\le \rho\|x_1-x_2\|, \forall (x_1, x_2)\in X_\varepsilon,
\end{equation*}
where $X_\varepsilon:=\{(x_1, x_2): \|x_1-x_2\|\le \underline\alpha^{-1}(\bar\alpha(\varepsilon))
+\underline\alpha^{-1}\big((\sigma_1(\varepsilon)+\max_{w\in \mathcal{W}}\{\sigma_2(\|w\|_\infty)\})
/{\mu}\big)+\varepsilon\}$, $\underline\alpha, \bar \alpha, \sigma_1, \sigma_2, \mu$ are defined in Definition \ref{def6}, $\mathcal{W}$ is the set of disturbance signals, and $\varepsilon$ is the desired precision.
\end{assumption}

Assumption \ref{ass1} is not conservative since it only requires Lipschitz continuity within a neighborhood of $x_1$, the radius of which is determined by the desired precision $\varepsilon$. Note that the Lipschitz constant $\rho$ is independent of $\varepsilon$. Then, we can get the following result.
%
%

\begin{theorem}\label{thm2}
Given the concrete system $\Sigma$ in (\ref{cs}) and the abstract system $\Sigma'$ in (\ref{abs}). Let $\varepsilon>0$ be the desired precision. Suppose Assumption \ref{ass1} holds. Assume that there exists a \emph{$\eta$-CGPS Lyapunov function} $V$ for $(\Sigma, \Sigma')$ and let $u_v$ be the associated control interface that is \emph{admissible}. If furthermore, one has that $\|w\|_{\infty } < \tilde\varepsilon:= \sigma_2^{-1}(\mu\underline\alpha(\bar \alpha^{-1}(\underline\alpha(\varepsilon/\rho)))/4), \forall w\in \mathcal{W}$; then, $\Sigma'\preceq_{\mathcal{S}}^{(\varepsilon, \tilde\varepsilon)} \Sigma$ if
\begin{equation}\label{thm2c2}
\begin{aligned}
  &\underline{\alpha}^{-1}(\bar \alpha(\eta))+\eta+\underline{\alpha}^{-1}\left(\bar{\alpha}\left(\underline{\alpha}^{-1}\left(\frac{4\sigma_1(\eta)}{\mu}\right)\right)\right)\\
  &\hspace{1cm}<\frac{\varepsilon}{\rho}-\underline{\alpha}^{-1}\left(\bar{\alpha}\left(\underline{\alpha}^{-1}\left(\frac{4\sigma_2(\|w\|_\infty)}{\mu}\right)\right)\right).
\end{aligned}
\end{equation}
\end{theorem}

\textbf{Proof:} By definition of $[\mathbb{R}^n]_\eta$, for all $x_0\in \mathbb{R}^n$, there exists $x'_0\in [\mathbb{R}^n]_\eta$ such that $\|x_0-x'_0\|\le \eta$. Then, one has from Assumption \ref{ass1} that
\begin{equation*}
\|h(x_0)-h(x'_0)\|\le \rho\|x_0-x'_0\|\le \varepsilon.
\end{equation*}
Hence, $(x_0, x'_0)\in \hat X_0$. Item\textit{ i)} of Definition \ref{def8} holds.

Given $(x_0, x'_0)\in \hat X_0$ and an input signal $v\in \mathcal{U}'$. Since the control interface $u_v$ is admissible, then one has $u(t)=u_v(t, v(t), \xi(x_0, u_v, w, t), \xi'(x'_0, v, t))\in U, \forall t\in \text{dom}(v)$. Thus, $u\in \mathcal{U}$.
Let $q(t)=\xi(x'_0, v, t))$. Then, one has $x_2(t)=\xi'(x'_0, v, t)=Q_\eta(q(t)), \forall t\in \text{dom}(v)$. Let also $x_1(t)=\xi(x_0, u_v, w, t))$, where $u_v$ is the admissible control interface. To prove item \textit{ii)} of Definition \ref{def8}, it is sufficient to prove that $\|h(x_1(t))-h(x_2(t)\|\le \varepsilon, \forall t\in \text{dom}(v)$.

Since $V$ is a $\eta$-CGPS Lyapunov function for $(\Sigma, \Sigma')$, then (\ref{lf3}) holds. One has from Theorem \ref{thm0} that
\begin{equation*}
\begin{aligned}
 \|x_1(t)-&q(t)\|\le \underline{\alpha}^{-1}(e^{-\frac{\mu}{2} t}\bar \alpha(\|x_1(0)-q(0)\|))\\
  &+\underline{\alpha}^{-1}(\bar{\alpha}(\underline{\alpha}^{-1}\left({4\sigma_1(\eta)}/{\mu}\right)))\\
  &+\underline{\alpha}^{-1}(\bar{\alpha}(\underline{\alpha}^{-1}\left({4\sigma_2(\|w\|_\infty))}/{\mu}\right))).
\end{aligned}
\end{equation*}
In addition, $\|x_1(0)-q(0)\|=\|\xi(0)-\xi'(0)\|=\|x_0-x'_0\|\le \eta$. Using (\ref{thm2c2}), one can further get
\begin{equation*}
\begin{aligned}
 \|x_1(t)- x_2(t)\|&\le \|x_1(t)-q(t)\|+\|q(t)-x_2(t)\|\\
 &= \|x_1(t)-q(t)\|+\|q(t)-Q_\eta(q(t))\|\\
 &\le {\varepsilon}/\rho,
\end{aligned}
\end{equation*}
and thus $\|h(x_1(t))- h(x_2(t))\|\le \rho\|x_1(t)- x_2(t)\|\le \varepsilon$. Item \textit{ii)} of Definition \ref{def8} holds and thus $\Sigma'\preceq_{\mathcal{S}}^{(\varepsilon, \tilde\varepsilon)} \Sigma$.
$\hspace*{\fill}\square$

\begin{remark}
  The construction of symbolic models and the implementation of the admissible control interface rely on the computation of the state-space abstraction and the abstract controller. For different systems, computational tools have been developed for this purpose, e.g., PESSOA \cite{PESSOA2009}, SCOTS \cite{Rungger2016}, and LTLCon \cite{Kloetzer2006}.
\end{remark}

\begin{remark}
One key step for the construction of symbolic models is to find an admissible control interface. From Definition \ref{def3}, one can see that for a control interface $u_v$ to be admissible, the key factor is to find an input map $U'$ admissible to $u_v$. When the input set for the concrete system is unbounded, i.e., $U=\mathbb{R}^m$, any control interface that maps to $\mathbb{R}^m$ is admissible. However, in practical applications, input saturations are common constraints. We note that when the input set for the concrete system is bounded, it is not always possible to find an admissible control interface. The good news is that, for a certain class of incrementally quadratic nonlinear systems, we show in the next section that it is possible to construct an admissible control interface $u_v$, such that $\Sigma$ robustly approximately simulates $\Sigma'$.
\end{remark}

\section{Incrementally quadratic nonlinear systems}\label{IQNS}

In this section, we consider a class of perturbed incrementally quadratic nonlinear systems  \cite{Dalto13}, for which the systematic construction of the admissible control interface and robust approximate symbolic models is possible. This kind of nonlinear systems are very useful and include many commonly encountered nonlinearities, such as the global Lipschitz nonlinearity, as special cases. In addition, many practical applications, such as vehicle models, manipulators, and electrical power convertors, are incrementally quadratic nonlinear systems.

Consider the nonlinear time-varying system described by
\begin{equation}\label{x2}
\Sigma_1: \left\{\begin{aligned}
{\dot x}(t) =& Ax(t) + Bu(t)+Ep(t, C_qx+D_qp)+w(t)\\
y(t)=&Cx(t),
\end{aligned}
\right.
\end{equation}
where $x\in \mathbb{R}^n, y\in \mathbb{R}^l$, $u\in U\subseteq \mathbb{R}^m$, and $w\in W\subset \mathbb{R}^n$ are the state, output, control input, and external disturbance, respectively, $p: \mathbb{R}_{\ge 0} \times \mathbb{R}^{l_p}\to \mathbb{R}^{l_e}$ represents the known continuous nonlinearity of the system, and $A, B, C, E, C_q, D_q$ are constants matrices of appropriate dimensions.

\begin{definition}\cite{Acikmese11}
Given a function $p: \mathbb{R}_{\ge 0} \times \mathbb{R}^{l_p}\to \mathbb{R}^{l_e}$, a symmetric matrix $M\in \mathbb{R}^{(l_p+l_e)\times (l_p+l_e)}$ is called an incremental
multiplier matrix for $p$ if it satisfies the following
incremental quadratic constraint for any $q_1, q_2\in \mathbb{R}^{l_p}$:
\begin{equation}\label{qc}
  {\left[ \begin{array}{l}
\;\;\;\;\;{q_2} - {q_1}\\
p(t, {q_2}) - p(t, {q_1})
\end{array} \right]^T}M\left[ \begin{array}{l}
\;\;\;\;\;{q_2} - {q_1}\\
p(t, {q_2}) - p(t, {q_1})
\end{array} \right] \ge 0.
\end{equation}
\end{definition}

\begin{remark}
The incremental quadratic constraint (\ref{qc}) includes a broad class
of nonlinearities as special cases. For instance, the globally
Lipschitz condition, the sector bounded nonlinearity, and the positive real nonlinearity $p^TSq\ge 0$ for some symmetric, invertible matrix $S$. Some other nonlinearities that can be expressed using the incremental quadratic constraint were discussed in \cite{Acikmese11,Dalto13}, such as the case when the Jacobian of
$p$ with respect to $q$ is confined in a polytope or a cone.
\end{remark}

\begin{assumption}\label{ass4}
There exist matrices $P=P^T\succ 0, L$ and a scalar $\alpha>0$ such that the following matrix inequality
\begin{equation}\label{lmi2}
\begin{aligned}
\left[ \begin{array}{l}
P(A+BL)+(A+BL)^TP+2\alpha P\;\;PE\\
\quad\quad\quad\quad\quad\quad{E^T}P\quad\quad\quad\quad\quad\quad\quad\quad 0
\end{array} \right]\\
 + {\left[\begin{array}{l}
{C_q}\;\;\;D_q\\
0\;\;\;\;\;I
\end{array} \right]^T}M\left[ \begin{array}{l}
{C_q}\;\;\;D_q\\
0\;\;\;\;\;I
\end{array} \right] \le 0
\end{aligned}
\end{equation}
is satisfied, where $M=M^T$ is an incremental
multiplier matrix for function $p$.
\end{assumption}

\begin{remark}
The matrix inequality (\ref{lmi2}) is not a LMI. Hence, one can not solve for $P, L$ reliably via, e.g., the interior point
method algorithms. However, we note that parameterization methods, such as block diagonal parameterization \cite{Acikmese11} can be utilized to transform (\ref{lmi2}) into Ricatti equations and/or LMIs under certain conditions. Moreover, we note that several necessary and/or sufficient conditions have been provided in \cite{Acikmese11,Dalto13} to guarantee the existence of solutions to  (\ref{lmi2}).
\end{remark}

The abstract system (obtained by applying the state-space discretization (\ref{sabs0})) is given by
\begin{equation}\label{xx2}
\Sigma'_1: \left\{\begin{aligned}
\xi(t)=&Q_\eta(\hat \xi(t))\\
\dot{\hat \xi}(t)=& A{\hat \xi}(t) + Bv(t)+Ep(t, C_q {\hat \xi}+D_qp),\\
\zeta(t)=&C\xi(t),
\end{aligned}
\right.
\end{equation}
where $v\in U'(t)$.

According to Remark \ref{rem1}, we first ignore the input constraint for the concrete system (\ref{x2}) by assuming that $U=\mathbb{R}^m$. The control interface $u_v: \mathbb{R}_{\ge 0}\times 2^{\mathbb{R}^m} \times \mathbb{R}^n \times [\mathbb{R}^n]_\eta \to \mathbb{R}^m$ is then designed as
\begin{equation}\label{u2}
u_v(t, v(t), x(t), \xi(t))=v(t)+L(x(t)-\xi(t)),
\end{equation}
where $L$ is the solution of (\ref{lmi2}).
One can verify that $u_v$ is admissible by letting $U'(t)=\mathbb{R}^m, \forall t\ge 0$.
Then, we get the following result.

\begin{theorem}\label{thm3}
Consider the concrete system (\ref{x2}) with the input set $U=\mathbb{R}^m$ and the abstract system (\ref{xx2}). Let $\varepsilon>0$ be the desired precision. The input $u(t)$ of (\ref{x2}) is synthesized by the control interface (\ref{u2}). Suppose that Assumption \ref{ass4} holds and the disturbance set $W$ satisfies $\|w\|_\infty<\tilde\varepsilon:=\alpha\varepsilon{\sqrt{\lambda _{\min }(P)}}/(2\|c\|\sqrt{{\lambda _{\max}}(P)}), \forall w\in \mathcal{W}$; then, $\Sigma'_1\preceq_{\mathcal{S}}^{(\varepsilon, \tilde\varepsilon)} \Sigma_1$ if the state-space discretization parameter $\eta$ satisfies
\begin{equation*}
\begin{aligned}
&\eta\le \left(\frac{\varepsilon}{\|C\|}-\frac{2\sqrt{{\lambda _{\max}}(P)}\|w\|_\infty}{\alpha\sqrt{{\lambda _{\min}}(P)}}\right)\\
&\hspace{1.5cm}\frac{\alpha\sqrt{{\lambda _{\min }(P)}}}{\alpha\sqrt{{\lambda _{\min }}(P)}+\sqrt{\alpha^2\lambda _{\max }(P)+2\|\hat L\|}},
\end{aligned}
\end{equation*}
where $\hat L=L^TB^TPBL$ and $P, L, \alpha$ are the solution to (\ref{lmi2}).
\end{theorem}

\textbf{Proof:} Let $e(t)=\xi(t)-\hat \xi(t)$, then one has $\|e(t)\|\le \eta, \forall t$. Define $\delta(t)=x(t)-\hat \xi(t)$. Then, from (\ref{x2}) and (\ref{xx2}) one has
\begin{equation*}
\begin{aligned}
\dot\delta(t)=&A\delta(t)+BL(\delta(t)+e(t))\\
&+E(p(t, C_q x+D_q p)-p(t, C_q\hat \xi+D_q p))+w(t)\\
=&A_c\delta(t)+BLe(t)+E\Phi_p(t, x, \hat \xi)+w(t),
\end{aligned}
\end{equation*}
where $A_c=A+BL$ and
\begin{equation*}
\Phi_p(t, x, \hat \xi)=p(t, C_qx+D_q p)-p(t, C_q\hat \xi+D_q p).
\end{equation*}
Post and pre multiplying both sides of inequality (\ref{lmi2}) by $(\delta(t), \Phi_p(t, x, \hat \xi))$ and its transpose and using condition (\ref{qc}) we obtain
\begin{equation*}
\delta^T P \dot{\delta}\le -\alpha \delta^T P\delta+\delta^T PBL e+\delta^T P w.
\end{equation*}
Consider the following Lyapunov function candidate
\begin{equation*}
V(t, x, \hat \xi)=(x-\hat \xi)^T P (x-\hat \xi).
\end{equation*}
Then, one has $\lambda_{\rm{min}}(P)\|x-\hat \xi\|^2\le V(t, x, \hat \xi)\le \lambda_{\rm{max}}(P)\|x-\hat \xi\|^2$.
Taking the derivative of $V$ on $t$, one has
\begin{equation}\label{dotV}
\begin{aligned}
\dot V(t, x, \hat \xi)&=2\delta^T P \dot{\delta}\\
&\le -2\alpha \delta^T P\delta+2\delta^T PBL e+2\delta^T P w\\
&\le -\alpha V(t, x, \hat \xi)+\frac{2}{\alpha}\|\hat L\|\eta^2+\frac{2}{\alpha}\|P\|\|w\|^2.
\end{aligned}
\end{equation}
Therefore, $V(t, x, \hat \xi)$ is a valid $\eta$-CGPS Lyapunov function for $(\Sigma_1, \Sigma'_1)$, where $\underline\alpha(x)=\lambda_{\rm min}(P)x^2, \bar\alpha(x)=\lambda_{\rm max}(P)x^2$, $\sigma_1(\eta)={2}\|\hat L\|\eta^2/{\alpha}$ and $\sigma_2 (\|w\|_{\infty })={2}\|P\|\|w\|_{\infty }^2/{\alpha}$. In addition, one can verify that Assumption \ref{ass1} holds with $\rho=\|C\|$. Then, the conclusion follows from Theorem \ref{thm2}.
$\hspace*{\fill}\square$

Next, we will show how to find an input map $U'$ admissible to $u_v$ when the real input set $U$ is considered.

From Theorem \ref{thm3}, we have (\ref{dotV}) holds. Then, using the comparison principle, we can further get
\begin{equation*}
\begin{aligned}
 &V(t, x(t), \hat \xi(t))\\
\le&e^{-\alpha t}V(t, x(0), \hat \xi(0))
+\frac{2\|\hat L\|\eta^2+2\|P\|\|w\|^2}{\alpha^2}(1-e^{-\alpha t})\\
\le&\lambda_{\max}(P)\eta^2+\frac{2\|\hat L\|\eta^2+2\|P\|\|w\|^2}{\alpha^2}.
\end{aligned}
\end{equation*}
Then, one can further have
\begin{equation*}
\begin{aligned}
\|x(t)-\hat \xi(t)\|\le \sqrt{\frac{V(t, x(t), \hat\xi(t))}{\lambda_{\min}(P)}}
\le K_1\eta+K_2\bar{w},
\end{aligned}
\end{equation*}
where $K_1=\sqrt{{\lambda_{\max}(P)}/{\lambda_{\min}(P)}+2{\|\hat L\|}/{(\alpha^2\lambda_{\min}(P))}}, \\ K_2=\sqrt{{2\lambda_{\max}(P)}/{(\alpha^2\lambda_{\min}(P))}}$, $\bar{w}=\max_{w\in \mathcal{W}}\{\|w\|_\infty\}$, and
\begin{equation*}
\begin{aligned}
\|x(t)-\xi(t)\|\le& \|x(t)-\hat \xi(t)\|+\|\hat \xi(t)-\xi(t)\|\\
\le &(K_1+1)\eta+K_2\bar{w}.
\end{aligned}
\end{equation*}
Define $e_u(t)=u(t)-v(t)$. Then, one has
\begin{equation}\label{euv}
\begin{aligned}
  \|e_u(t)\|&=\|L(x(t)-\xi(t))\|\\
 & \le \|L\|((K_1+1)\eta+K_2\bar{w}).
\end{aligned}
\end{equation}
From (\ref{euv}), one can see that the norm of the relative error between $u(t)$ and $v(t)$, i.e., $\|e_u(t)\|$ is upper bounded, and the radius of the upper bound is determined by $\eta$ and $\bar{w}$ (due to the special form of control interface that was designed in (\ref{u2})). Let
\begin{equation*}
  \tilde U=\big\{z\in U| {\rm d}(z, F_r(U))<
  \|L\|((K_1+1)\eta+K_2\bar{w})\big\},
\end{equation*}
be the set of points in $U$, whose distance to the boundary of $U$ is less than $\|L\|((K_1+1)\eta+K_2\bar{w})$. Then, by choosing
\begin{equation*}
  U'(t)=U\setminus\tilde U, \forall t\ge 0,
\end{equation*}
one can guarantee that $u(t)\in U, \forall v(t)\in U'(t), \forall t\ge 0$. Moreover, we note that when $\Sigma_1$ is deterministic, i.e., $w(t)\equiv 0$, one can always find $U'(t)\neq \emptyset, \forall t\ge 0$ for all $U, \rm{int{U}}\neq \emptyset$ since $U'(t)\to U$ as $\eta\to 0$.

\section{Simulation}

In this section, two simulation examples are provided to validate the effectiveness of the theoretical results.

\subsection{Example 1}

Consider the (undisturbed) time-varying nonlinear system $\Sigma$ given by
\begin{equation}\label{csex2}
  \Sigma:\left\{\begin{aligned}
  \dot x_1(t)&= Ax_1(t)+\frac{1}{t+1}\sin(x_1(t))+u(t)\\
  y_1(t)&=x_1(t),
  \end{aligned}\right.
\end{equation}
where $x_1, y_1, u\in \mathbb{R}^2$, $A=[0.15, 0;0, 0.05]$
is a constant matrix and the sinusoidal function $\sin(\cdot)$ is defined element-wise. One can verify that (\ref{csex2}) is not $\delta$-ISS.

Applying the state abstraction (\ref{sabs0}), then the abstract system can be written as
\begin{equation}\label{asex2}
  \Sigma':\left\{\begin{aligned}
  x_2(t)&= Q_\eta(\hat x_2(t))\\
  \dot{\hat x}_2(t)&= A\hat x_2(t)+\sin(\hat x_2(t))/(t+1)+v(t)\\
  y_2(t)&=x_2(t),
  \end{aligned}\right.
\end{equation}
where $v(t)\in U'(t)$. Since $U=\mathbb{R}^2$, one can choose the input map $U'$ as $U'(t)=\mathbb{R}^2, \forall t\ge 0$, which is admissible to any $u_v$.

The control interface $u_v: \mathbb{R}_{\ge 0} \times \mathbb{R}^n\times \mathbb{R}^n \times [\mathbb{R}^n]_\eta \to \mathbb{R}^n$ is designed as
\begin{equation}\label{uex2}
\begin{aligned}
  u_v(t, v(t), x_1(t), x_2(t))=v(t)+P^{-1}R(x_1(t)-x_2(t)),
\end{aligned}
\end{equation}
where $P=I_2, R=-5.4I_2$ are the solution to the following LMI
\begin{equation*}
\begin{aligned}
\left[ \begin{array}{l}
A^TP+PA+2R+2\alpha P\;\;P\\
\quad\quad\quad\quad\quad P\quad\quad\quad\quad\quad 0
\end{array} \right]+\left[ \begin{array}{l}
nI_n\;\;\;\;0_n\\
\;\;0_n\;\;-I_n
\end{array} \right]
 \le 0
\end{aligned}
\end{equation*}
with a scalar $\alpha=3.7$. Let $\varepsilon=0.5$ be the desired precision. According to Theorem \ref{thm3}, the desired precision $\varepsilon=0.5$ can be achieved by choosing the state-space discretization parameter $\eta=0.18$. The goal is to stabilize the system $\Sigma$ to a unit ball around the origin.

\begin{figure}[h!]
\centering
\includegraphics[width=.45\textwidth]{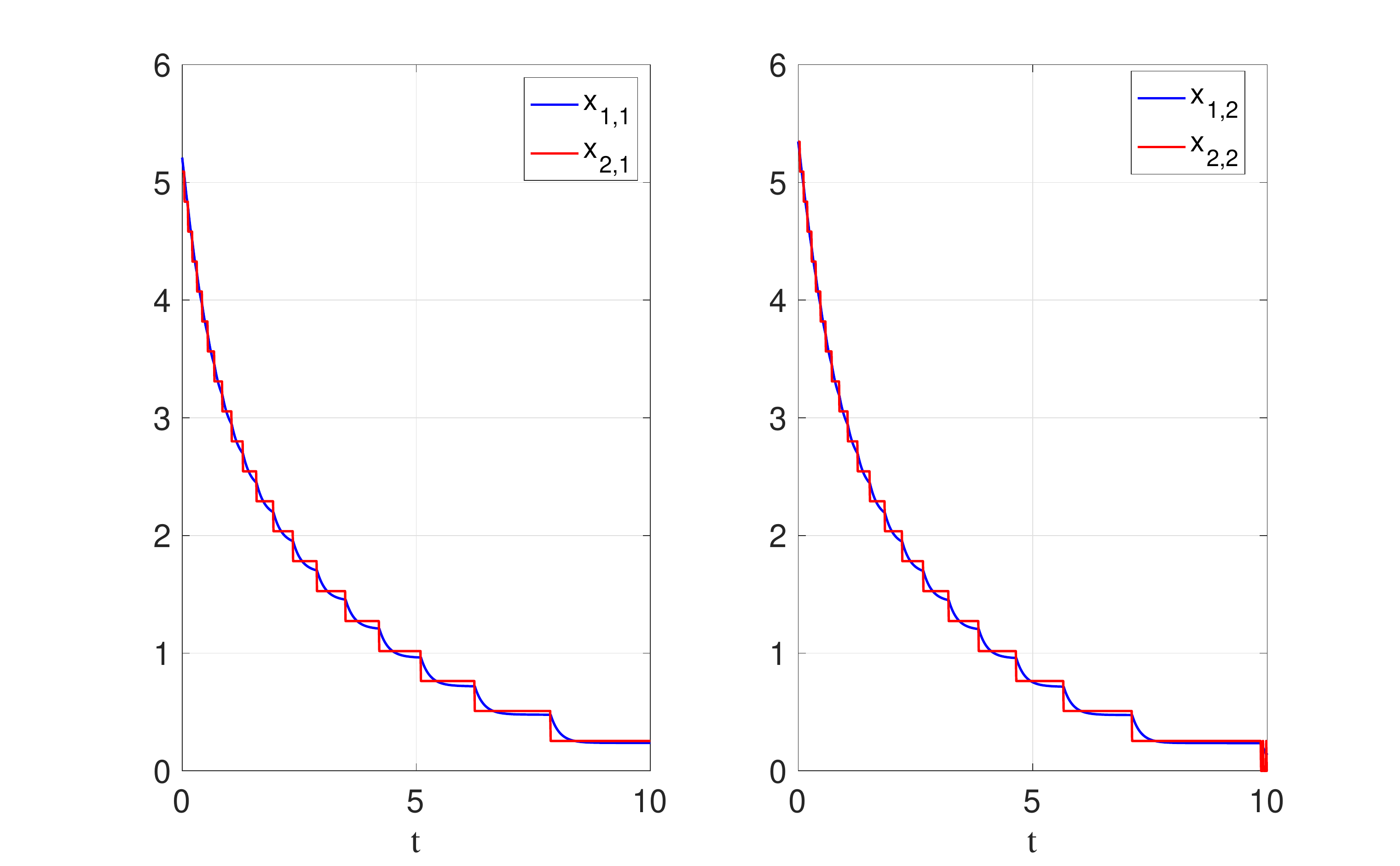}
\caption{The output trajectories of the concrete system $\Sigma$ (blue line) and the abstract system $\Sigma'$ (red line). }\label{fig1}
\end{figure}

\begin{figure}[h!]
\centering
\includegraphics[width=.45\textwidth]{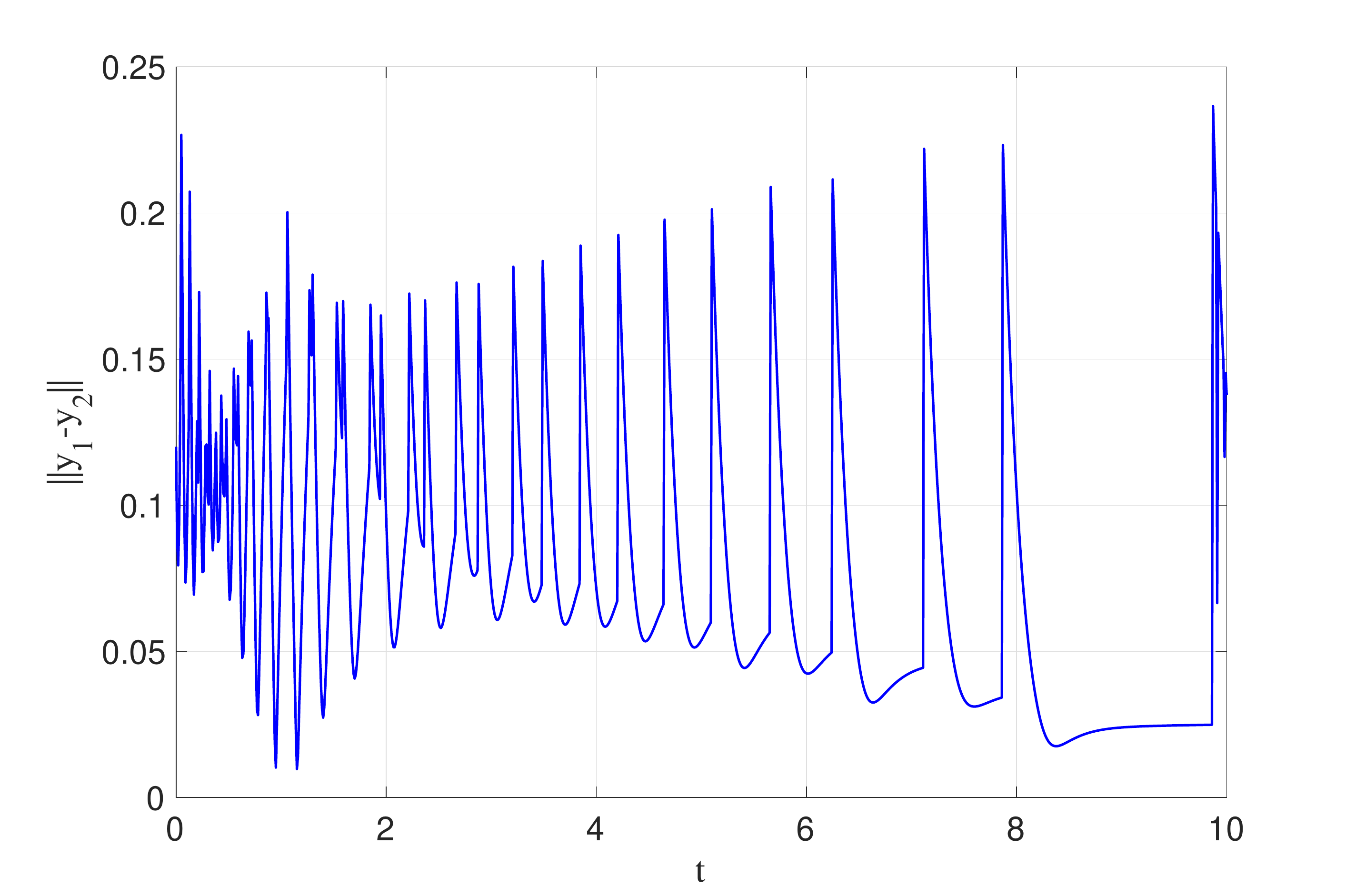}
\caption{The evolution of the output error $\|y_1-y_2\|$.}\label{fig2}
\end{figure}

The simulation results are shown in Figs. \ref{fig1}-\ref{fig2}. The trajectory $x_2$ of $\Sigma'$ is obtained by applying a stabilization controller $v(t)$, and it is represented by the
red line in Fig. \ref{fig1} ($x_{2,1}, x_{2,2}$ are the two state components of $x_2$). The trajectory $x_1$ of $\Sigma$ is obtained via the control interface (\ref{uex2}), and it is represented by the  blue line in Fig. \ref{fig1} ($x_{1,1}, x_{1,2}$ are the two state components of $x_1$). The evolution of the output error $\|y_1-y_2\|$ is depicted in Fig. \ref{fig2}, and one can see that the desired precision 0.5 is satisfied at all times.


In this example, the desired precision is $\varepsilon=0.5$ while the simulation result in Fig. \ref{fig2} shows that the
output error $\|y_1-y_2\|$ is at most 0.25. This means that the theoretical bound of $\eta$ obtained using Theorem \ref{thm3} can be rather conservative (due to the use of Lyapunov-like function).

%

\subsection{Example 2}
\begin{figure}[h!]
\centering
\includegraphics[width=.45\textwidth]{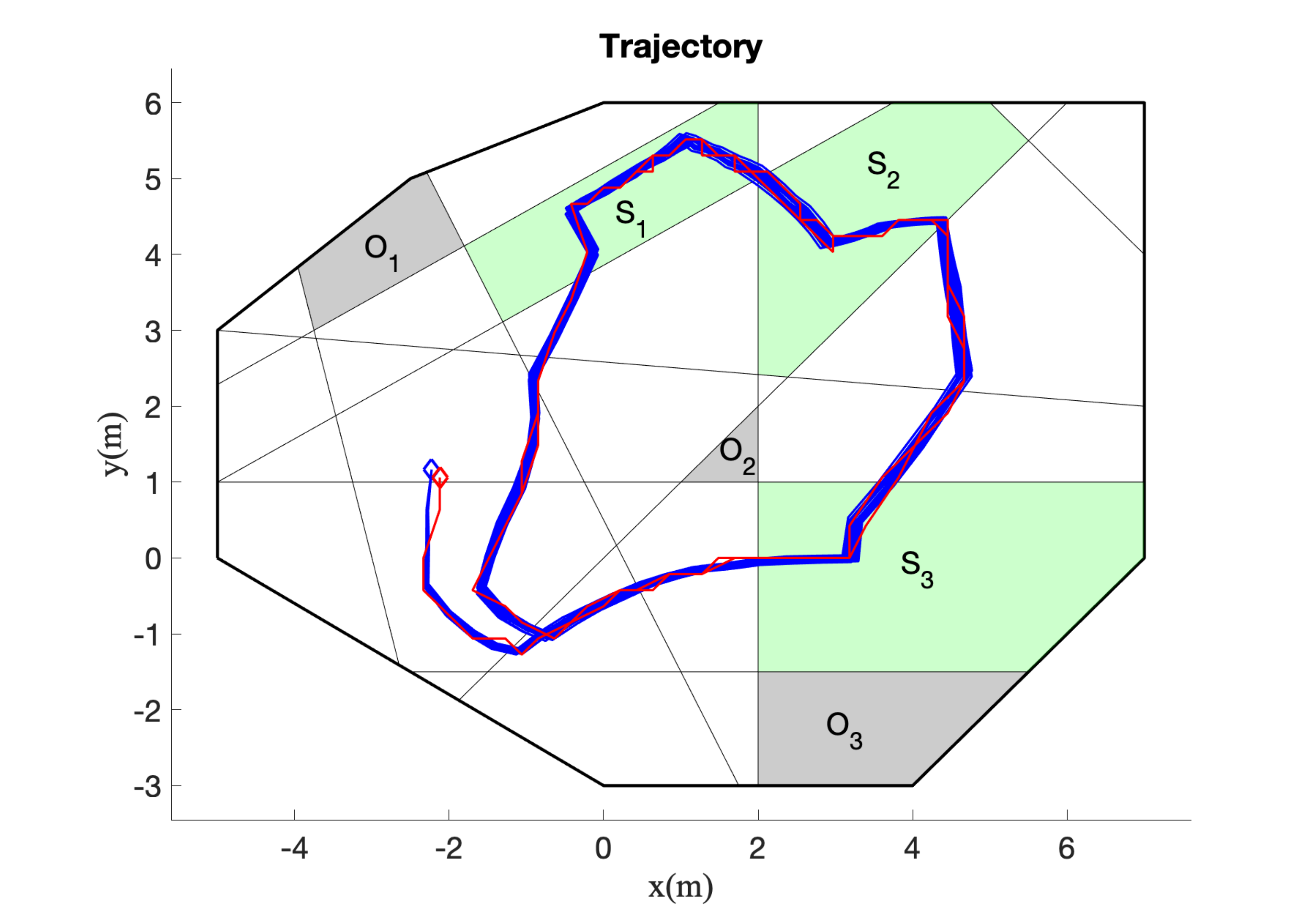}
\caption{Output trajectories of the concrete system $\Sigma_2$ (blue lines) for 100 realizations of disturbance signals and output trajectory of the abstract system $\Sigma'_2$ (red line). }\label{fig4}
\end{figure}

Consider a mobile robot moving in $\mathbb{R}^2$, the dynamics of which is given by:
\begin{equation}\label{sim_ex}
  \Sigma_2: \left\{\begin{aligned}
  \dot x_1&=Ax_1+Bu+w\\
       y_1&=x_1,
  \end{aligned}\right.
\end{equation}
where
\begin{equation*}
  A=\left[ \begin{array}{l}
0.2\quad 0.3\\
0.5\quad -0.5
\end{array} \right], B=\left[ \begin{array}{l}
1 \quad 0\\
0 \quad 1
\end{array} \right].
\end{equation*}
The input set $U=[-5, 5]\times [-5, 5]$ and the disturbance set $W=[-0.05, 0.05]\times [-0.05, 0.05]$. The problem is to drive the robot in the bounded workspace $\mathbb{W}$ shown in Fig. \ref{fig4}, where the three grey solid polygons $O_1, O_2, O_3$ represent obstacles and the three green solid polygons $S_1, S_2, S_3$ represent target regions. The goal of the motion planning problem consists in visiting all the three target regions $S_1, S_2, S_3$ infinitely many times while avoiding collision with the obstacles. This specification can be represented by a linear temporal logic (LTL) \cite{Baier2008} formula $\phi=\mathsf{G}\mathbb{W} \wedge \mathsf{G}(\neg (O_1\vee O_2 \vee O_3)) \wedge \mathsf{G} (\mathsf{F}S_1\wedge \mathsf{F}S_2 \wedge \mathsf{F}S_3)$, where $\neg, \wedge, \vee$ are negation, logic `AND', logic `OR' operators, respectively, and $\mathsf{G}, \mathsf{F}$ are temporal operators `ALWAYS' and `EVENTUALLY', respectively. The details about the syntax and semantics of LTL can be found in \cite{Baier2008}, Chapter 5.

Let the desired precision be $\varepsilon=1$. The control interface is designed as
\begin{equation}\label{u}
u_v(t, v(t), x(t), \xi(t))=v(t)-\frac{1}{2}B^TP(x(t)-\xi(t)),
\end{equation}
where $P$ is the solution to the ARE
\begin{equation*}
{A^T}P + PA -PB{B^T}P + I_2 = 0.
\end{equation*}
According to Theorem \ref{thm3}, the desired precision $\varepsilon$ can be achieved by choosing the state-space discretization parameter $\eta=0.15$.  Then, by further choosing $U'(t)=[-3.5, 3.5]\times [-3.5, 3.5], \forall t\ge 0$, one can guarantee that the control interface (\ref{u}) is admissible. The abstract system (obtained by applying the state-space abstraction (\ref{sabs0})) is denoted by $\Sigma'_2$ and the output of $\Sigma'_2$ is denoted by $y_2$.

Using the LTL control synthesis toolbox LTLCon \cite{Kloetzer2006}, we first synthesize a trajectory and the associated control policy for the abstract system $\Sigma'_2$, which is shown by the red solid line in Fig. \ref{fig4}. One can see that any trajectory remaining within the distance 1 from this trajectory satisfies the problem specification.

\begin{figure}[h!]
\centering
\includegraphics[width=.45\textwidth]{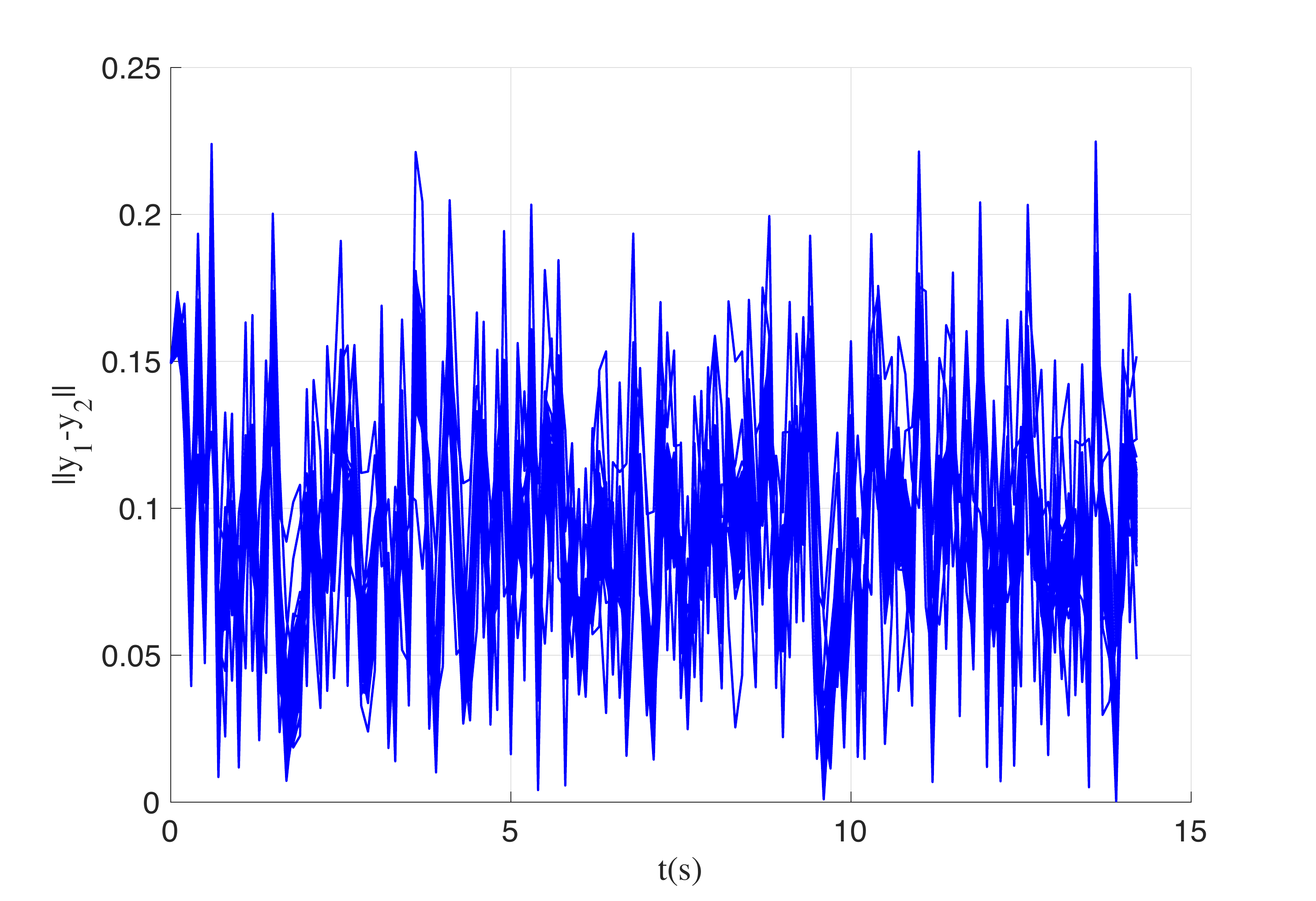}
\caption{The evolution of $\|y_1-y_2\|$ for 100 realizations of disturbance signals.}\label{fig5}
\end{figure}

\begin{figure}[h!]
\centering
\includegraphics[width=.45\textwidth]{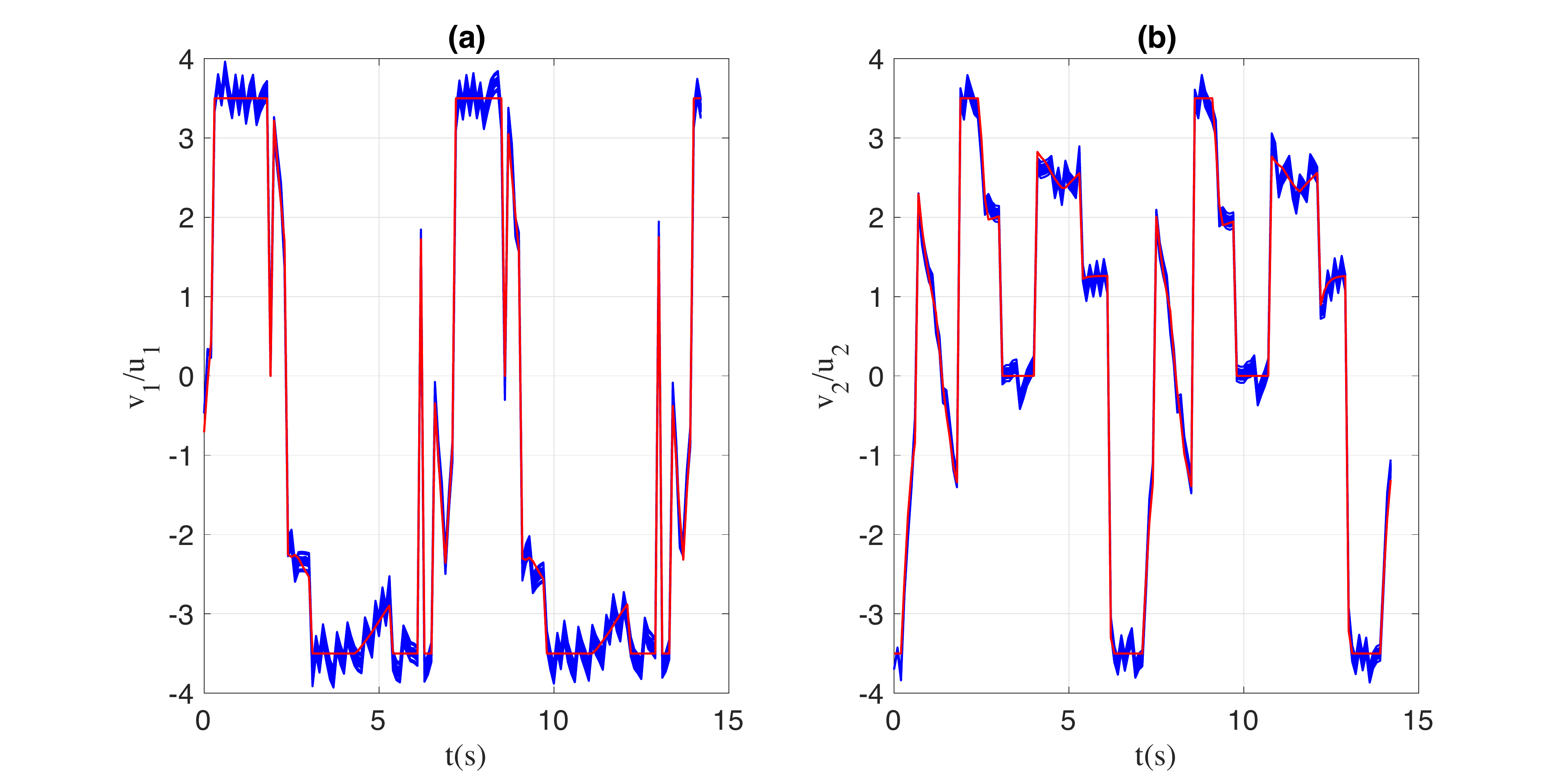}
\caption{The evolution of the inputs $u$ (blue lines) for 100 realizations of disturbance signals and $v$ (red line).}\label{fig6}
\end{figure}

The output trajectory $y_1$ of $\Sigma_2$ is obtained by applying the synthesized input for the abstract system $\Sigma'_2$ via the control interface (\ref{u}). Furthermore, in order to validate robustness, we run 100 realizations of the disturbance trajectories. The
resulting trajectories for these 100 realizations are shown (by the solid blue line) in Fig. \ref{fig4}.
One can see that all the trajectories satisfy the goal of the motion planning problem. The evolution of the output error $\|y_1-y_2\|$ for the 100 realizations is depicted in Fig. \ref{fig5}, and one can see that the desired precision is preserved at all times. In addition, the evolution of the input components $v_1, v_2$ for the abstract system $\Sigma'_2$ and the input components $u_1, u_2$ for the concrete system $\Sigma_2$ are plotted in Fig. \ref{fig6}, respectively. One can see that $u\in U$ (i.e., the input constraint is satisfied) at all times.

Similar to Example 1, the desired precision is $\varepsilon=1$ in this example while the simulation result in Fig. \ref{fig5} shows that the output
error $\|y_1-y_2\|$ is at most 0.25. This again means that the theoretical bound of $\eta$ can be conservative.

\section{Conclusion}

This paper involved the construction of discrete state-space symbolic models for continuous-time uncertain nonlinear systems. Firstly, a stability notion called $\eta$-CGPS and its Lyapunov function characterizations were proposed. After that, a notion of robust approximate simulation relation was further introduced. It was shown that every continuous-time uncertain concrete system, under the condition that there exists an admissible control interface such that the augmented system can be made $\eta$-CGPS, robustly approximately simulates its discrete state-space abstraction. In the future, more efficient abstraction techniques, such as multi-scale abstraction \cite{Gossler15}, will be taken into account and experimental validation will be pursued.

\medskip

\begin{thebibliography}{0}

\bibitem{Acikmese11}
A\c{c}{\i}kme\c{s}e, B., \& Corless, M. (2011). Observers for systems with nonlinearities satisfying incremental quadratic constraints. \emph{Automatica}, 47(7), 1339-1348.

\bibitem{Alur00}
Alur, R., Henzinger, T. A., Lafferriere, G., \& Pappas, G. J. (2000). Discrete abstractions of hybrid systems. \emph{Proceedings of the IEEE}, 88(7), 971-984.

\bibitem{Angeli99}
Angeli, D., \& Sontag, E. D. (1999). Forward completeness, unboundedness observability, and their Lyapunov characterizations. \emph{Systems \& Control Letters}, 38(4-5), 209-217.

\bibitem{Angeli02}
Angeli, D. (2002). A Lyapunov approach to incremental stability properties. \emph{IEEE Transactions on Automatic Control}, 47(3), 410-421.

\bibitem{Arnold03}
Arnold, A., Vincent, A., \& Walukiewicz, I. (2003). Games for synthesis of controllers with partial observation. \emph{Theoretical Computer Science}, 28(1), 7–34.

\bibitem{Baier2008}
Baier, C., \& Katoen, J. P. (2008). Principles of model checking. MIT press.

\bibitem{Cassandras99}
Cassandras, C., \& Lafortune, S. (1999). Introduction to Discrete Event Systems. Boston, MA: Kluwer.

\bibitem{Dalto13}
D'Alto, L., \& Corless, M. (2013). Incremental quadratic stability. Numerical Algebra, \emph{Control \& Optimization}, 3(1), 175-201.

\bibitem{Fu13}
Fu, J., Shah, S., \& Tanner, H. G. (2013, June). Hierarchical control via approximate simulation and feedback linearization. \emph{In 2013 American Control Conference (pp. 1816-1821)}. IEEE.

\bibitem{Girard_Pappas07}
Girard, A., \& Pappas, G. J. (2007). Approximation metrics for discrete and continuous systems. \emph{IEEE Transactions on Automatic Control}, 52(5), 782-798.

\bibitem{Girard_Pappas09}
Girard, A., \& Pappas, G. J. (2009). Hierarchical control system design using approximate simulation. \emph{Automatica}, 45(2), 566-571.

\bibitem{Girard_Pola10}
Girard, A., Pola, G., \& Tabuada, P. (2009). Approximately bisimilar symbolic models for incrementally stable switched systems. \emph{IEEE Transactions on Automatic Control}, 55(1), 116-126.

\bibitem{Girard12}
Girard, A. (2012). Controller synthesis for safety and reachability via approximate bisimulation. \emph{Automatica}, 48(5), 947-953.

\bibitem{Gossler15}
Girard, A., G\"{o}ssler, G., \& Mouelhi, S. (2015). Safety controller synthesis for incrementally stable switched systems using multiscale symbolic models. \emph{IEEE Transactions on Automatic Control}, 61(6), 1537-1549.

\bibitem{Goedel12}
Goedel, R., Sanfelice, R. G., \& Teel, A. R. (2012). Hybrid dynamical systems: modeling stability, and robustness. \emph{Princeton University Press}.

\bibitem{Khail02}
Khalil, H. K. (2002). Nonlinear Systems. \emph{Prentice Hall}, 3rd edn.

\bibitem{Kim2017}
Kim, E. S., Arcak, M., \& Seshia, S. A. (2017). Symbolic control design for monotone systems with directed specifications. \emph{Automatica}, 83, 10-19.

\bibitem{Kloetzer2006}
Kloetzer, M., \& Belta, C. (2008). A fully automated framework for control of linear systems from temporal logic specifications. \emph{IEEE Transactions on Automatic Control}, 53(1), 287-297.

\bibitem{Kumar95}
Kumar, R., \& Garg, V. (1995). Modeling Control of Logical Discrete Event Systems. Boston, MA: Kluwer.

\bibitem{Liu2016}
Liu, J., \& Ozay, N. (2016). Finite abstractions with robustness margins for temporal logic-based control synthesis. \emph{Nonlinear Analysis: Hybrid Systems}, 22, 1-15.

\bibitem{Madhusudan03}
Madhusudan, P., Nam, W., \& Alur, R. (2003). Symbolic computational techniques for solving games. \emph{Electronic Notes in Theoretical Computer Science}, 89(4), 578-592.

\bibitem{Mallik2018}
Mallik, K., Schmuck, A. K., Soudjani, S., \& Majumdar, R. (2018). Compositional synthesis of finite-state abstractions. \emph{IEEE Transactions on Automatic Control}, 64(6), 2629-2636.

\bibitem{Milner89}
Milner, R. (1989). Communication and concurrency. \emph{Prentice Hall}.

\bibitem{Park81}
Park, D. (1981). Concurrency and automata on infinite sequences. \emph{In Theoretical computer science (pp. 167-183)}. Springer, Berlin, Heidelberg.

\bibitem{PESSOA2009}
Mazo, M., Davitian, A., \& Tabuada, P. (2010, July). Pessoa: A tool for embedded controller synthesis. \emph{In International Conference on Computer Aided Verification} (pp. 566-569). Springer, Berlin, Heidelberg.

\bibitem{Ramadge87}
Ramadge, P. J., \& Wonham, W. M. (1987). Modular feedback logic for discrete event
systems. \emph{SIAM Journal on Control and Optimization}, 25(5), 1202-1218.

\bibitem{Reissig17}
Reissig, G., Weber, A., \& Rungger, M. (2016). Feedback refinement relations for the synthesis of symbolic controllers. \emph{IEEE Transactions on Automatic Control}, 62(4), 1781-1796.

\bibitem{Rungger2016}
Rungger, M., \& Zamani, M. (2016). SCOTS: A tool for the synthesis of symbolic controllers. \emph{In Proceedings of the 19th international conference on hybrid systems: Computation and control (pp. 99-104)}.

\bibitem{Saoud2020}
Saoud, A., Girard, A., \& Fribourg, L. (2020). Contract-based design of symbolic controllers for safety in distributed multiperiodic sampled-data systems. \emph{IEEE Transactions on Automatic Control}, DOI: 10.1109/TAC.2020.2992446.

\bibitem{Smith18(2)}
Smith, S. W., Arcak, M., \& Zamani, M. (2018). Approximate abstractions of control systems with an application to aggregation. \emph{arXiv preprint arXiv:1809.03621}.

\bibitem{Tabuada09}
Tabuada, P. (2009). Verification and control of hybrid systems: a symbolic approach. \emph{Springer Science \& Business Media}.

\bibitem{Yang17}
Yang, K., \& Ji, H. (2017). Hierarchical analysis of large-scale control systems via vector simulation function. \emph{Systems \& Control Letters}, 102, 74-80.

\bibitem{Pian19}
Yu, P., \& Dimarogonas, D. V. (2019). Approximately symbolic models for a class of continuous-time nonlinear systems. \emph{arXiv preprint arXiv:1909.09040}.

\bibitem{Zamani12}
Zamani, M., Pola, G., Mazo, M., \& Tabuada, P. (2011). Symbolic models for nonlinear control systems without stability assumptions. \emph{IEEE Transactions on Automatic Control}, 57(7), 1804-1809.

\bibitem{Zamami13}
Zamani, M., van de Wouw, N., \& Majumdar, R. (2013). Backstepping controller synthesis and characterizations of incremental stability. \emph{Systems \& Control Letters}, 62(10), 949-962.

\bibitem{Zamani14}
Zamani, M., Esfahani, P. M., Majumdar, R., Abate, A., \& Lygeros, J. (2014). Symbolic control of stochastic systems via approximately bisimilar finite abstractions. \emph{IEEE Transactions on Automatic Control}, 59(12), 3135-3150.


\end{thebibliography}

\end{document}